\documentclass[aps,prd,onecolumn,amssymb]{revtex4}
\usepackage{graphicx,bm,color}
\usepackage[english]{babel}
\usepackage{amsmath}
\usepackage{amsfonts}
\begin{document}

\newcommand{\be}{\begin{equation}}
\newcommand{\ee}{\end{equation}}
\newcommand{\bea}{\begin{eqnarray}}
\newcommand{\eea}{\end{eqnarray}}
\newcommand{\nn}{\nonumber \\}
\newcommand{\e}{\mathrm{e}}

\title{Modified Gauss-Bonnet  gravity with Lagrange multiplier constraint as mimetic theory}

\author{Artyom~V.~Astashenok$^{1}$,
Sergei~D.~Odintsov$^{2,3}$, V.K. Oikonomou$^{4,5}$}
\affiliation{$^1$ Baltic Federal University of I. Kant, Department
of
Theoretical Physics, 236041, 14, Nevsky st., Kaliningrad, Russia \\
$^2$Instituci\`{o} Catalana de Recerca i Estudis Avan\c{c}ats
(ICREA),
Barcelona, Spain \\
$^3$Institut de Ciencies de l'Espai (CSIC-IEEC), Campus UAB,
Campus UAB, Carrer de Can Magrans, s/n 08193 Cerdanyola del Valles, Barcelona, Spain\\
$^{4)}$ Department of Theoretical Physics, Aristotle University of
Thessaloniki,
54124 Thessaloniki, Greece\\
$^{5)}$  National Research Tomsk State University,  634050 Tomsk and
Tomsk State Pedagogical University,  634061 Tomsk, Russia }

\begin{abstract}

In this paper we propose and extensively study mimetic $f({\cal G})$
modified gravity models, with various scenarios of cosmological
evolution, with or without extra matter fluids. The easiest
formulation is based on the use of Lagrange multiplier constraint.
 In certain versions of this
theory, it is possible to realize accelerated expansion of the
Universe or even unified evolution which includes inflation with
dark energy, and at the same time in the same theoretical framework,
dark matter is described by the theory. This is achieved by the
re-parametrization of the metric tensor, which introduces a new
degree of freedom in the cosmological equations and leads to
appearance of the mimetic ``dark matter'' component. In the context
of mimetic $f({\cal G})$ theory, we also provide some quite general
reconstruction schemes, which enable us to find which $f({\cal G})$
gravity generates a specific cosmological evolution.  In addition,
we also provide the general reconstruction technique for Lagrange
multiplier $f({\cal G})$ gravity. All our results are accompanied by
illustrative examples, with special emphasis on bouncing
cosmologies.
\end{abstract}

\maketitle

\section{Introduction}

The recent observational data \cite{planck,bicep}, have increased
the scientific interest in cosmology, since early-time acceleration
is favored by present observations. In addition to this early-time
acceleration, in the late 90's, the late-time acceleration
(including that from modified gravity \cite{Capozziello2002,
Carroll2004}) was also verified by the observations
\cite{latetimeobse}. Since then a great task for cosmologists is to
consistently describe these two acceleration eras, and if possible,
to include them in the same theoretical framework as it was first
proposed in \cite{NO1}. Modified gravity \cite{reviews1}, is one
appealing candidate for a complete cosmological theory, in the
context of which late-time acceleration can be consistently
described but more importantly, the unification of early-time
acceleration with late-time acceleration \cite{NO1,odintsov2003} can
be achieved (for a general review, see \cite{reviews1}). There are
many types of modified theories of gravity with higher-derivative
terms, such as $f(R)$ modified gravity \cite{FR}, $f({\cal{G}})$
theories of gravity \cite{other}, etc. In this paper, we shall
introduce a new type of mimetic $f({\cal{G}})$ theory of gravity,
which as we shall demonstrate, can successfully describe a variety
of cosmological scenarios. It is inspired by mimetic dark matter
models \cite{mukhanov1,mukhanov2,Golovnev} and serves as a
generalization of both $f({\cal{G}})$ and mimetic dark matter
models.

The mimetic dark matter models \cite{mukhanov1,mukhanov2,Golovnev},
are generalizations of general relativity, at which general
relativity is modified in a minimal way, so that the conformal
symmetry not to be violated. In the theoretical formalism of mimetic
dark matter theories, an extra internal degree of freedom appears,
which can describe dark matter via such mimetic gravity theory
\cite{mukhanov1} (see also Refs.\cite{mukhanov2,Golovnev}). In order
for the conformal symmetry not to be violated, the metric is not
regarded as the fundamental variable of this gravitational theory,
but is decomposed into a new metric and a scalar field, with the
latter being the extra gravitational degree of freedom which
actually describes dark matter.

A recent study in which the formalism of mimetic gravity was used in
the context of $f(R)$ gravity was done in Ref. \cite{NO2}. As was
shown in that work, in the context of mimetic $f(R)$ gravity and
specifically of modified $f(R)$ gravity with Lagrange multiplier
constraint, it is possible to realize the unification of early-time
and late-time acceleration eras as well as having Dark Matter in
unified scenario.

It is of great importance to understand how universal is such an
approach to the unification of the three dark sectors of the
Universe.

In this paper we consider modified Gauss-Bonnet  $f({\cal G})$
gravity, which was introduced as gravitational alternative to dark
energy in Ref. \cite{NO3}, but generalized in such a way that the
Lagrangian multiplier constraint is used, as done in Ref.
\cite{CMO}. The specific choice of the Lagrange multiplier
constraint, renders such a theory as a mimetic modified Gauss-Bonnet
gravity of certain type. The accelerating cosmology generated by
such a mimetic gravity, is studied in detail and we also investigate
which $f({\cal G})$ gravity can realize various cosmological
scenarios, with special emphasis to bouncing cosmologies
\cite{bounce1,bounce2,bounce3,bounce4,bounce5,superbounce}. We need to note that the idea of obtaining a bouncing solution in the frame of Lagrange multiplier modified gravity was studied for the first time in \cite{caisar}, which is very relevant in spirit to our approach. In
addition, we also introduce the mimetic gravity formalism, in the
context of a general Jordan frame $f({\cal G})$ gravity. As a
general remark, we must note that within the context of mimetic
$f({\cal G})$ gravity and Lagrange multiplier modified $f({\cal G})$
gravity, it is possible that there exists a large number of $f({\cal
G})$ theories which can consistently describe a given cosmological
evolution.

This paper is organized as follows: In section II we present the
theoretical formalism of mimetic Gauss-Bonnet gravity. The most
important argument is based on the fact that the metric is
redefined, and all cosmological equations are obtained by varying
the redefined action with respect to redefined metric
$\hat{g}_{\mu\nu}=-\hat{g}^{\rho\sigma}\partial_{\rho}\phi
\partial_{\sigma}\phi \hat{g}_{\mu\nu}$ and mimetic scalar field
$\phi$ instead of $g_{\mu\nu}$. In section III we investigate how
various cosmological scenarios can be realized in the context of
mimetic $f({\cal G})$ gravity, with special emphasis to cosmological
bounces. In section IV, we consider a generalization of $f({\cal
G})$ theory, based on the introduction of a Lagrange multiplier,
which includes the mimetic case. Several examples are worked out and
also comparison to the standard Einstein-Hilbert gravity with
Lagrange multiplier, worked out in \cite{mukhanov2} is performed.
Moreover, we also compare the resulting modified $f({\cal G})$
gravity, mimetic or Lagrange multiplier modified, with the standard
$f({\cal G})$ gravity picture. The conclusions follow in the end of
the paper.

\section{Mimetic $f({\cal G})$ gravity}

Consider a general $f({\cal G})$ gravity, with the Jordan frame
action being equal to,
\begin{equation}
\label{action} S=\int d^4x\sqrt{-g}\left[\frac{R}{2\kappa^2}+f({\cal
G})\right]+S_m.
\end{equation}
where $S_m$ stands for the action of matter-fluids, which induces
the energy momentum tensor $T_{\mu\nu}$ in the field equations (for
some incomplete list of works on cosmology of such theory, see
\cite{other}). We adopt the standard geometrical conventions of
Einstein-Hilbert gravity, with regards to the commutative connection
of the underlying Riemannian spacetime, $R$ is the Ricci scalar, and
the function $f({\cal G})$ corresponds to a generic globally
differentiable function of the Gauss-Bonnet topological invariant
${\cal G}$, which is equal to,
\begin{equation}
{\cal
G}=R^2-4R_{\mu\nu}R^{\mu\nu}+R_{\mu\nu\lambda\sigma}R^{\mu\nu\lambda\sigma}\,,
\end{equation}
where $R_{\mu\nu}$ and $R_{\mu\nu\lambda\sigma}$ are the Ricci  and
Riemann tensors, respectively. We adopted the signature for the
Riemannian metric as $(-+++)$ and also we assume that $\kappa^2=8\pi
G/c^4=1$, where $G$ is the standard Newtonian gravitational
coupling.

In order to obtain the mimetic $f({\cal G})$, the following
parametrization of metric is assumed
\cite{mukhanov1,mukhanov2,Golovnev}:
\begin{equation}\label{100}
g_{\mu\nu}=-\hat{g}^{\rho\sigma}\partial_{\rho}\phi
\partial_{\sigma}\phi \hat{g}_{\mu\nu}.
\end{equation}
By varying the metric we obtain the following relation,
$$
\delta g_{\mu\nu}=\hat{g}^{\rho\tau}\delta\hat{g}_{\tau\omega}
\hat{g}^{\omega\sigma}\partial_{\rho}\phi \partial_{\sigma}\phi
\hat{g}_{\mu\nu} -\hat{g}^{\rho\sigma}\partial_{\rho}\phi
\partial_{\sigma}\phi \delta\hat{g}_{\mu\nu}-
$$
$$
-2\hat{g}^{\rho\sigma}\partial_{\rho}\phi
\partial_{\sigma}\delta\phi \hat{g}_{\mu\nu}.
$$
In addition, by varying the action (\ref{action}), with respect to
the redefined metric $\hat{g}_{\mu\nu}$, instead of the standard
Jordan frame metric $g_{\mu\nu}$, and with respect to the mimetic
scalar $\phi$, the resulting field equations take the following
form,
\begin{equation}\label{eom}
R_{\mu\nu}- \frac{1}{2}R g_{\mu\nu}+
\end{equation}
$$
+8\Big[R_{\mu\rho\nu\sigma}+R_{\rho\nu}g_{\sigma\mu}-R_{\rho\sigma}g_{\nu\mu}-R_{\mu\nu}g_{\sigma\rho}+R_{\mu\sigma}g_{\nu\rho}
+\frac{R}{2}\left(g_{\mu\nu}g_{\sigma\rho}-g_{\mu\sigma}g_{\nu\rho}\right)\Big]\nabla^{\rho}\nabla^{\sigma}f_{\cal
G}+ \left(f_{\cal G}{\cal G}-f({\cal G})\right)g_{\mu\nu}+
$$
$$
+\partial_{\mu}\phi\partial_{\nu}\phi\left(-R+8\left(-R_{\rho\sigma}
+\frac{1}{2}R
g_{\rho\sigma}\right)\nabla^{\rho}\nabla^{\sigma}f_{\cal
G}+4(f_{\cal G}{\cal G}-f({\cal G}))\right)=
$$
$$
=T_{\mu\nu}+\partial_{\mu}\phi\partial_{\nu}\phi T,
$$
where $f_{\cal G}$ stands for $f_{\cal G}=d f({\cal G}) / d{ \cal
G}$. We also note that the covariant derivative $\nabla_{\mu}$, acts
on vectors as
$\nabla_{\mu}V_{\nu}=\partial_{\mu}V_{\nu}-\Gamma_{\mu\nu}^{\lambda}V_{\lambda}$
and accordingly it acts on the metric $g_{\mu\nu}$.

By varying the action (\ref{action}) with respect to the scalar
field $\phi$, we obtain the following equation,
\begin{equation}\label{FG}
\nabla^{\mu}\left(\partial_{\mu}\phi\left(-R+8\left(-R_{\rho\sigma}+\frac{1}{2}
R g_{\rho\sigma}\right)\nabla^{\rho}\nabla^{\sigma}f_{\cal
G}+4(f_{\cal G}{\cal G}-f({\cal G})\right)-T)\right)=0.
\end{equation}
Assuming a spatially-flat Friedmann-Robertson-Walker (FRW) metric,
with metric \be ds^{2}=-dt^{2}+a^{2}(t)(dx^{2}+dy^{2}+dz^{2}) \ee
the scalar curvature and the Gauss-Bonnet invariant have the
following form:
$$R=6\left(\frac{\ddot{a}}{a}+\frac{\dot{a}^2}{a^2}\right)=6(\dot{H}+2H^2),$$

$${\cal G}=-24\frac{\ddot{a}\dot{a}^2}{a^3}=-24H^2(\dot{H}+H^2). \label{gbinvariant}$$
where $H(t)$ denotes as usual the Hubble rate
$H(t)=\dot{a}(t)/a(t)$. From Eq. (\ref{100}) we find that
\begin{equation}\label{introeqns}
g^{\mu\nu}\partial_{\mu}\phi\partial_{\nu}\phi=-1
\end{equation}
and since the scalar field $\phi$ depends only on the cosmic time,
this results to the constraint $\phi=t$. Taking into account the
aforementioned constraint, we can obtain the $(t,t)$ component of
the expression given in (\ref{eom}):
\begin{equation}\label{FRIED-1}
2\dot{H}+3H^{2}+16H(\dot{H}+H^{2})\frac{d f_{{\cal
G}}}{dt}+8H^{2}\frac{d^2 f_{{\cal G}} }{dt^2} -(f_{{\cal G}}{\cal
G}-f({\cal G}))=-p.
\end{equation}
The same equation results if we consider the $(r,r)$ component. By
integrating Eq. (\ref{FG}), we obtain the following relation,
\begin{equation}
-R+8\left(-R_{\rho\sigma}+\frac{1}{2}R
g_{\rho\sigma}\right)\nabla^{\rho}\nabla^{\sigma}f_{{\cal
G}}+4(f_{{\cal G}}{\cal G}-f({\cal G}))+\rho-3p=-\frac{C}{a^3}
\end{equation}
and finally this results to,
\begin{equation}\label{FRIED-2}
\dot{H}+2H^{2}+4H^{2}\frac{d^{2}f_{{\cal G}}}{dt^2}+
4H\left(2\dot{H}+3H^{2}\right)\frac{d f_{{\cal
G}}}{dt}+\frac{2}{3}(f_{{\cal G}}{\cal G}-f({\cal G}))+
\frac{\rho}{6}-\frac{p}{2}=-\frac{C}{a^3}.
\end{equation}
where $C$ is an arbitrary constant.

Combining Eqs. (\ref{FRIED-1}) and (\ref{FRIED-2}), we obtain:
\begin{equation}\label{fried3}
\dot{H}+4H^{2}\frac{d^{2}f_{{\cal
G}}}{dt^2}+4H(2\dot{H}-H^{2})\frac{df_{{\cal
G}}}{dt}=-\frac{1}{2}\left(\rho+p\right)-\frac{C}{a^{3}}.
\end{equation}
where for simplicity we redefined the constant $C$. It is convenient
to introduce the function $g(t)$, which is defined to be equal to,
$$
g(t)=\frac{df_{{\cal G}}}{dt},
$$
which satisfies the equation:
\begin{equation}\label{FRIED4}
4H^{2}\frac{dg(t)}{dt}+4H(2\dot{H}-H^2)g(t)=-\dot{H}-\frac{1}{2}\left(\rho+p\right)-\frac{C}{a^{3}}.
\end{equation}
For $C=0$ Eqs. (\ref{FRIED-1}) and (\ref{fried3}) coincide with the
Friedmann equations in ordinary $f({\cal G})$ gravity. If the right
hand side of Eq. (\ref{FRIED4} is zero, that is,
$B(t)=-\dot{H}-\frac{1}{2}\left(\rho+p\right)-\frac{C}{a^{3}}= 0$
then the solution to this equation can be written as follows,
\begin{equation}
g(t)=g_{0}\left(\frac{H_{0}}{H}\right)^{2}\exp\left(\int_{{0}}^{t}Hdt\right).
\end{equation}
In the above equation, $g_{0}$ is a constant and $H_{0}=H({0})$. In
the case that the right hand of Eq. (\ref{FRIED4}) is non-zero, that
is when $B(t)\neq 0$, the solution $g(t)$ reads,
\begin{equation}
g(t)=g_{0}\left(\frac{H_{0}}{H}\right)^{2}\exp\left(\int_{{0}}^{t}Hdt\right)+\frac{1}{4H^{2}}\int_{0}^{t}dt_{1}B(t_{1})\exp\left(\int_{t_{1}}^{t}
H(\tau)d\tau\right).\label{genericbouncesol}
\end{equation}
Therefore one can specify the cosmic evolution of the Universe, in
terms of the Hubble rate $H(t)$ and then obtain the function $g(t)$.
Then by integrating this, with respect to the cosmic time, we can
obtain the function $f(t)$, that is,
$$
f_{{\cal G}}(t)=\int g(t) dt.
$$
Using the explicit dependence of the Gauss-Bonnet invariant on the
cosmic time $t$, one can get the explicit form of the function
$t({\cal G})$ and therefore find the function $f_{{\cal G}}({\cal
G})$. Then, upon integrating with respect to ${\cal{G}}$, we easily
obtain the mimetic $f({\cal G})$ gravity. In the next section, we
shall present some illustrative examples, in order to demonstrate
explicitly how the reconstruction method works, and also in order to
see the differences between Jordan frame mimetic $f({\cal G})$
gravity and ordinary Jordan frame $f({\cal G})$ gravity.

\section{Cosmological models with various evolutions from mimetic $f({\cal G})$}

\subsection{A general example}

We start off with the cosmological evolution of the Universe, with a
Hubble rate equal to,
$$\label{apopano}
H(t)=\frac{H_{0}}{1+\alpha t}, \quad \alpha>0
$$
which could correspond for example to early-time acceleration from
the moment $t=0$ to some later time $t$, with $t>>\alpha^{-1}$. The
corresponding scale factor for the Hubble parameter (\ref{apopano})
is equal to,
\begin{equation}
a(t)=a_{0}\left(1+\alpha t\right)^{H_{0}/\alpha}
\end{equation}
The corresponding Gauss-Bonnet invariant is,
$${\cal G}=24\frac{H_{0}^{2}(\alpha-H_{0})}{(1+\alpha
t)^{3}}.\label{wdt}
$$
The solution of Eq. (\ref{FRIED4}) with zero right hand side
($B=0$), is equal to,
$$
g(t)=g_{0}\left(1+\alpha t\right)^{H_{0}/\alpha+2}.
$$
For an empty universe without matter and mimetic dark matter
($\rho=p=C=0$) the function $g(t)$ is equal to,
$$
g(t)=g_{0}\left(1+\alpha
t\right)^{H_{0}/\alpha+2}-\frac{1}{4H_{0}}\frac{1+\alpha
t}{\alpha+H_{0}}.
$$
Then, the resulting expression for the function $f_{{\cal G}}(t)$ is
equal to,
$$
f_{{\cal G}}(t)=f_{0}\left(1+\alpha
t\right)^{H_{0}/\alpha+3}-\frac{1}{8\alpha H_{0}}\frac{(1+\alpha
t)^{2}}{\alpha+H_{0}}.
$$
Using the Gauss-Bonnet invariant (\ref{wdt}), we can express the
function $f_{{\cal G}}(t)$, as a function of the Gauss-Bonnet
invariant ${\cal G} $,
$$
f_{{\cal G}}=A{\cal G}^{-1-H_{0}/3\alpha}+D{\cal G}^{-2/3},
$$
$$
D=-\frac{1}{8\alpha
H_{0}(\alpha+H_{0})}\left(24H_{0}^{2}(\alpha-H_{0})\right)^{2/3}.
$$
Finally, integrating the above equation, with respect to the
Gauss-Bonnet invariant ${\cal{G}}$, we obtain the resulting
non-mimetic $f({\cal G})$ gravity,
\begin{equation}\label{nomtc}
f({\cal G})=F_{0} {\cal G}^{-H_{0}/\alpha}+3D{\cal G}^{1/3}.
\end{equation}
For a Universe filled with matter and mimetic dark matter, which
means that $C\neq 0$, $\rho= \rho_{0} a^{-3}$, $p=0$, the
corresponding solution $g(t)$ has the following form,
\begin{equation}\label{axris}
g(t)=g_{0}\left(1+\alpha
t\right)^{H_{0}/\alpha+2}-\frac{1}{4H_{0}}\frac{1+\alpha
t}{\alpha+H_{0}}+\frac{\tilde{C}}{4H_{0}-\alpha}\frac{1}{4H_{0}^{2}}\left(1+\alpha
t\right)^{3(1-H_{0}/\alpha)}, \quad \tilde{C}=C+\rho_{0},
\end{equation}
and the corresponding solution $f_{{\cal G}}(t)$ reads,
\begin{equation}
f_{{\cal G}}(t)=A{\cal G}^{-1-H_{0}/3\alpha}+D{\cal G}^{-2/3}+E{\cal
G}^{-4/3+H_{0}/\alpha},
\end{equation}
where $E$ stands for,
$$
E=\frac{\tilde{C}}{4H_{0}^{2}}\frac{\alpha}{(4H_{0}-\alpha)(4\alpha-H_{0})}\left(24H_{0}^{2}(\alpha-H_{0})\right)^{4/3-H_{0}/\alpha},
$$
The resulting mimetic $f({\cal G})$ gravity easily follows,
\begin{equation}\label{answ}
f({\cal G})=F_{0} {\cal G}^{-H_{0}/\alpha}+3D{\cal
G}^{1/3}+\frac{3\alpha}{3H_{0}-\alpha}E{\cal G}^{-1/3+H_{0}/\alpha}.
\end{equation}
By looking Eqs. (\ref{answ}) and (\ref{nomtc}), we can see the
difference between non-mimetic and mimetic $f({\cal G})$ gravity.
Notice that in the mimetic gravity case, a given cosmological
evolution can be realized by a family of $f({\cal G})$ functions.
For example by putting $C=-\rho_{0}$ and therefore the last term can
be omitted in Eq. (\ref{axris}). It is worthy to note that in
ordinary $f({\cal G})$ gravity with matter fluids, we have no such
freedom in choosing the $f({\cal G})$ function, and a unique
solution for $f({\cal G})$ exists (in (\ref{axris}) one need to
change $\tilde{C}\rightarrow \rho_{0}$).

Let us study another case, in which the cosmological evolution is
described by the following Hubble rate,
$$
H=\frac{H_{0}}{1+\beta^2 t^{2}}
$$
We consider the simple case for which $g_{0}=0$ and $\rho=p=C=0$,
and the solution $g(t)$ is equal to,
\begin{equation}
g(t)=-\frac{\beta^{2}(1+\beta^2 t^2)(1+H_{0}t-\beta^2
t^2)}{2H_{0}(H_{0}^{2}+4\beta^2)}
\end{equation}
The Gauss-Bonnet invariant is,
$$
{\cal G}=\frac{24H_{0}^{2}(2\beta^2 t-H_{0})}{(\beta^2
t^{2}+1)^{3}}.
$$
For $t\rightarrow \infty$ we have the following asymptotic behavior:
$$
g(t)\thickapprox
\frac{\beta^{6}t^{4}}{2H_{0}(H_{0}^{2}+4\beta^{2})},
$$
$$
{\cal G}\thickapprox \frac{48H_{0}^{2}}{\beta^{4}t^{5}}.
$$
Therefore, the function $f_{{\cal G}}({\cal G})$ is equal to,
$$
f_{{\cal G}}({\cal G})\thickapprox
\frac{\beta^{14/5}}{2(48H_{0})^{-4/5}H_{0}(H_{0}^{2}+4\beta^{2})}{\cal
G}^{-4/5}
$$
and upon integrating with respect to ${\cal G}$, we finally obtain
the $f({{\cal G}})$ gravity,
\begin{equation}
f({{\cal G}})\thickapprox
\frac{5\beta^{14/5}}{2(48H_{0})^{-4/5}H_{0}(H_{0}^{2}+4\beta^{2})}{\cal
G}^{1/5}.
\end{equation}

\subsection{Reconstruction of mimetic $f({\cal G})$ models from
bounce cosmological evolution}

Using the method we described in the previous sections, it is
possible to reconstruct an $f({\cal G})$ model for given
cosmological evolution with scale factor $a(t)$. In this section we
shall consider bouncing cosmologies, and we investigate which
$f({\cal G})$ model can generate such a cosmology, in the presence
of matter and mimetic dark matter.

\subsubsection{Superbounce from mimetic $f({\cal G})$ gravity}

In the context of $f({\cal G})$ modified gravity, various bouncing
Universe scenarios can be materialized straightforwardly, which in
the context of Einstein-Hilbert general relativity could only be
realized if the null energy condition is violated for a period of
time. For reviews and an important stream of papers on bouncing
cosmology see \cite{bounce1,bounce2,bounce3,bounce4,bounce5} and
references therein. A bounce evolution has the following
characteristics: The Universe starts with a contraction up to a
point where a non-zero minimal radius is reached, where the scale
factor is non-vanishing too, and then the Universe starts to expand
again. In this way the singularity problem of the scale factor,
which is known as the initial singularity, is resolved in a
self-consistent way. Therefore, bouncing cosmology serves as a
convenient way to avoid singularities of crushing type, where the
geodesics incompleteness occurs. Let the bounce occur at some time
$t_s$, then for $t=t_s$ the Hubble rate is zero, that is $H(t_s)=0$,
while during the expanding phase, $H(t)>0$, and during the
contracting phase, $H(t)<0$.

Therefore the bouncing cosmology is one appealing candidate for the
consistent description of the Universe's evolution. Within this
context, we shall make use of two bouncing cosmologies for our
reconstruction examples.

We start of with the superbounce model \cite{superbounce}, which has
the following scale factor,
\begin{equation}\label{scsupe}
a(t)=(-t+t_s)^{\frac{2}{c^2}}
\end{equation}
while the corresponding Hubble rate is,
\begin{equation}\label{hu}
H(t)=-\frac{2}{c^2 (-t+t_s)}
\end{equation}
Using Eqs. (\ref{scsupe}) and (\ref{hu}) and substituting to Eq.
(\ref{genericbouncesol}), the solution $g(t)$ easily follows,
\begin{equation}\label{generic1}
g(t)=\frac{1}{4} c^4 g_0 H_0^2 (t-t_s)^{\frac{2}{c^2}}
(-t+t_s)^2-\frac{1}{32} c^6 (t-t_s)^{\frac{2}{c^2}} (-t+t_s)^2
\left(\frac{2}{-2+c^2}+\frac{c^2 C (-t+t_s)^{2-\frac{6}{c^2}}}{8-6
c^2+c^4}+\frac{c^2 (t-t_s)^2 (p+\rho )}{2 \left(2+3
c^2+c^4\right)}\right)
\end{equation}
The Gauss-Bonnet invariant for the Hubble rate of Eq. (\ref{hu}) is
equal to,
\begin{equation}\label{gabne}
{\cal G}=-96 c^2 t^2 (2 c + 4 c^2 t^2)
\end{equation}
and by solving Eq. (\ref{gabne}) with respect to $t$ we obtain two
real solutions,
\begin{equation}\label{g1}
t_1=\frac{-2 \sqrt{2} 3^{1/4} c^4 \sqrt{G} \left(c^8
\left(-2+c^2\right) G\right)^{1/4}+c^8 G t_s}{c^8
G},{\,}{\,}{\,}t_2=\frac{2 \sqrt{2} 3^{1/4} c^4 \sqrt{G} \left(c^8
\left(-2+c^2\right) G\right)^{1/4}+c^8 G t_s}{c^8 G}
\end{equation}
By substituting Eq. (\ref{g1}) to Eq. (\ref{generic1}) and
integrating over $\cal G$, we obtain two different functional forms
of mimetic $f ({\cal G})$ gravity, which are of the following form,
\begin{align}\label{genericfr}
& f ({\cal G})=-12 \left(-2+c^2\right)
(t({\cal{G}})-t_s)^{-1+\frac{2}{c^2}} \Big{(}\frac{4 \left(8 g_0
H_0^2+c^2 \left(1-4 g_0
H_0^2\right)\right)}{\left(-2+c^2\right)^2 \left(2+3 c^2\right)} \\
\notag & -\frac{2 c^4 C
(-t({\cal{G}})+t_s)^{2-\frac{6}{c^2}}}{\left(-4+c^2\right)^2
\left(8-14 c^2+5 c^4\right)}-\frac{c^4 (t({\cal{G}})-t_s)^2 (p+\rho
)}{\left(2+c^2\right)^2 \left(2+7 c^2+5 c^4\right)}\Big{)}
\end{align}
where $t({\cal{G}})=t_{1,2}$ with $t_1,t_2$ being defined in Eq.
(\ref{g1}).

Let us here compare the Jordan frame mimetic $f ({\cal G})$ modified
gravity solutions we found, that generate the superbounce
(\ref{scsupe}), with the solutions of non-mimetic $f ({\cal G})$
gravity in the Jordan frame. The non-mimetic $f ({\cal G})$ gravity
solutions that generate the superbounce (\ref{scsupe}) are the
following \cite{superbounce}:
\begin{equation}\label{finalsuperbouncefgs}
f ({\cal G})=\frac{t_*^2 {\cal G}-4 \sqrt{(-11+a) a^3 {\cal G}}}{4
a+4 a^2},{\,}{\,}{\,}f ({\cal G})=\frac{t_*^2 {\cal G}-2
\sqrt{(-11+a) a^3 {\cal G}} (1+{\cal G})}{4 a (1+a) {\cal G}}
\end{equation}
where $a=2/c^2$. Comparing Eqs. (\ref{genericfr}) and
(\ref{finalsuperbouncefgs}), we can see the obvious functional
differences. But there is also another important difference, having
to do with the freedom in choosing the solutions in the case of
mimetic $f ({\cal G})$ gravity, an issue we already discussed
earlier. Notice also that there is a similarity between the mimetic
and non-mimetic $f ({\cal G})$ gravities, which is that in both
cases, the resulting $f ({\cal G})$ gravity consists of two distinct
solutions, when the superbounce cosmological evolution is taken into
account.

\subsubsection{Generic bounce scenario from mimetic $f({\cal G})$
gravity}

Another interesting bouncing cosmology scenario is described by the
following scale factor,
\begin{equation}\label{sc1}
a(t)=e^{c t^2}
\end{equation}
with $c$ some arbitrary constant and the scale factor is normalized
again so that at the bouncing point, which is $t=0$, the scale
factor is equal to $a(0)=1$. The corresponding Hubble rate is
particularly simple, and it is equal to,
\begin{equation}\label{hubra1}
H(t)=2 c t
\end{equation}
Using the Hubble rate (\ref{hubra1}) and by substituting to Eq.
(\ref{genericbouncesol}), we get the function $g(t)$,
\begin{align}\label{bigone}
& g(t)=\frac{C e^{-2 c t^2}}{64 c^3 t}-\frac{e^{c t^2}}{16 c^2
t}-\frac{e^{c t^2} g_0 H_0^2}{4 c^2 t}-\frac{e^{c t^2} p}{64 c^3
t}-\frac{e^{c t^2} \rho }{64 c^3 t}+\frac{C \sqrt{\frac{\pi }{2}}
\text{erf}\left[\sqrt{2} \sqrt{c} t\right]}{32
c^{5/2}}+\frac{\sqrt{\pi } \text{erfi}\left(\sqrt{c} t\right)}{16
c^{3/2}} \\ \notag & +\frac{g_0 H_0^2 \sqrt{\pi }
\text{erfi}\left(\sqrt{c} t\right)}{4 c^{3/2}}+\frac{p \sqrt{\pi }
\text{erfi}\left(\sqrt{c} t\right)}{64 c^{5/2}}+\frac{\sqrt{\pi }
\rho  \text{erfi}\left(\sqrt{c} t\right)}{64 c^{5/2}} \\
\notag & -\frac{C t {\,}
 _2F_2(\frac{1}{2};\frac{3}{2};ct^2)}{16 c^2}-\frac{t {\,}
 _2F_2(\frac{1}{2};\frac{3}{2};ct^2)}{8
c}-\frac{p t{\,}
 _2F_2(\frac{1}{2};\frac{3}{2};ct^2)}{32 c^2}-\frac{t \rho
 _2F_2(\frac{1}{2};\frac{3}{2};ct^2)}{32 c^2}
\end{align}
where the functions $\text{erfi}(x)$ and $\text{erf}(x)$ denote the
imaginary error function and the error function respectively, while
the function $ _pF_q(a;b;z)$ stands for the generalized
hypergeometric function. The Gauss-Bonnet invariant in this case is
equal to,
\begin{equation}\label{generalizgbinv}
{\cal G}=-96 c^2 t^2 \left(2 c+4 c^2 t^2\right)
\end{equation}
which can be solved with respect to $t$, and we have the following
positive solution,
\begin{equation}\label{inser}
t=\sqrt{-\frac{1}{4 c}+\frac{\sqrt{24 c^6-c^4 G}}{8 \sqrt{6} c^4}}
\end{equation}
Having this at hand, and by substituting to Eq. (\ref{bigone}), we
easily obtain the function $f_{{\cal G}}({\cal G})$. Then by
integrating with respect to $\cal {G}$, we can get the resulting
$f_{\cal{G}}(\cal {G})$ function, but we omit the resulting
expression, since it's functional form is too large and complicated.
Note however that all the integrals can be evaluated analytically,
as the reader may convince himself.

Before closing, we study here a final example, for which we assume
that $g_{0}=0$, in which case the function $g(t)$ is equal to,
\begin{equation}
g(t)=-\frac{a^{3}(t)}{4\dot{a}^{2}(t)}\int_{0}^{t}a^{-1}(t_{1})\left(\frac{\ddot{a}(t_{1})}{a(t_{1})}-\frac{\dot{a}(t_{1})}{a^{2}(t_{1})}\right)dt_{1}-
\frac{\tilde{C}a^{3}(t)}{8\dot{a}^{2}(t)}\int_{0}^{t}a^{-4}(t_{1})dt_{1}.
\end{equation}
Consider a cosmological evolution with scale factor $a(t)=(1+\alpha
t)^{\gamma}$. The case $\alpha<0$, $\gamma<0$ corresponds to the Big
Rip singularity \cite{BigRip}, which occurs at $t_{s}=1/\alpha$. The
accelerated expansion corresponds to $\alpha>0$, $\gamma>1$. The
corresponding solution for $g(t)$ reads,
\begin{equation}\label{SolG}
g(t)=-\frac{1}{4\gamma^2\alpha^2}\left((1+\alpha
t)^{2+\gamma}-(1+\alpha t)^2\right)
-\frac{\tilde{C}}{8\gamma^{2}\alpha^{3}(4\gamma-1)}\left((1+\alpha
t)^{2+\gamma}-(1+\alpha t)^{3(1-\gamma)}\right), \quad \gamma\neq
1/4.
\end{equation}
The corresponding Gauss-Bonnet invariant ${\cal G}$ is,
$$
{\cal G}=24\frac{\gamma^{3}\alpha^{4}(1-\gamma)}{(1+\alpha t)^{4}}
$$
so the resulting expression for the $f({\cal G})$ gravity reads,
\begin{equation}
f({\cal G})=A{\cal G}^{1/4}+ B{\cal G}^{(1-\gamma)/4}+D{\cal
G}^{3\gamma/4}.
\end{equation}

%%%%%%%%%%%%%%%%%%%%%%%%%%%%%%%%%%%%%%%%%%%%%%%%%%%%%%%%%%%%%%%%%%%%%%%%%%%%%%%%

\section{$f({\cal G})$ gravity in formulation with Lagrange multiplier}

In this section we consider the formulation of $f({\cal G})$
modified gravity by using a Lagrange multiplier, and the impact of
this multiplier to the cosmological evolution of various models. One
can impose a condition on the scalar field by adding to the standard
action for $f({\cal G})$ gravity the following term,
$$\lambda
(g^{\mu\nu}\partial_{\mu}\phi\partial_{\nu}\phi+U(\phi)),$$ where
$\lambda$ is the Lagrange multiplier and $U(\phi)$ is a function of
the scalar field, in  principle arbitrarily chosen. We also include
in the Jordan frame action of the $f({\cal G})$ gravity, the action
of a scalar field (canonical or phantom) $\phi$ with arbitrarily
chosen scalar potential $V(\phi)$, which is of the form:
$$
S_{\phi}=\int d^{4}x\sqrt{-g}\left(-\epsilon
g^{\mu\nu}\partial_{\mu}\phi\partial_{\nu}\phi-V(\phi)\right).
$$
The cases $\epsilon={\pm 1}$ describe canonical and phantom scalar
field respectively. For $\epsilon=0$ the conformal symmetry of the
system is respected and the corresponding gravity with Lagrange
multiplier is equivalent to the mimetic $f({\cal G})$ gravity
considered above (for $U=1$ and $V=0$). (The case with non-zero $V$
maybe called as extended mimetic theory). The full Jordan frame
action of the $f({\cal G})$ gravity with Lagrange multiplier and in
the presence of the scalar field and other matter fluids, is the
following,
\begin{equation}\label{laction}
S=\int d^{4}x\sqrt{-g}\left(\frac{R}{2}+f({\cal G})-\epsilon
g^{\mu\nu}\partial_{\mu}\phi\partial_{\nu}\phi-V(\phi)+\lambda(g^{\mu\nu}\partial_{\mu}\phi\partial_{\nu}\phi+U(\phi))+{\cal
L}_{m}\right)
\end{equation}
By varying this action with respect to the metric $g_{\mu\nu}$, we
obtain the following two equations of motion:
\begin{equation}\label{FGFR-1}
3H^2+24H^{3}\frac{df_{{\cal G}}({\cal G})}{dt}+f({\cal G})-f_{{\cal
G}}({\cal G}){\cal
G}=\rho+\epsilon\dot{\phi}^{2}+V(\phi)-\lambda(\dot{\phi}^{2}+U(\phi)),
\end{equation}

\begin{equation}\label{FGFR-2}
-2\dot{H}-3H^2-8H^2\frac{d^{2}f_{{\cal G}}({\cal
G})}{dt^{2}}-16H(\dot{H}+H^2)\frac{df_{{\cal G}}({\cal
G})}{dt}+f_{{\cal G}}({\cal G}){\cal G}-f({\cal
G})=p+\epsilon\dot{\phi}^{2}-V(\phi)-\lambda(\dot{\phi}^{2}-U(\phi)).
\end{equation}
where we assumed a flat FRW metric. Also, by varying action
(\ref{laction}), with respect to the Lagrange multiplier, provides
us with the following constraint for the scalar field:
\begin{equation}
\dot{\phi}^{2}-U(\phi)=0.
\end{equation}
Finally for the scalar field we have the equation of motion:
\begin{equation}\label{SF}
2\partial_{t}((\lambda-\epsilon)
\dot{\phi})+6H(\lambda-\epsilon)\dot{\phi}-V'(\phi)+\lambda
U'(\phi)=0.
\end{equation}
For a Universe dominated by collisionless matter ($p=0$), the scalar
potential $V(t)$ and the Lagrange multiplier function $\lambda(t)$,
as function of the cosmic time read,
\begin{equation}\label{p1}
V(t)=2\dot{H}+3H^2+\epsilon\dot{\phi}^{2}+8H^2\frac{d^{2}f_{{\cal
G}}({\cal G})}{dt^{2}}+16H(\dot{H}+H^2)\frac{df_{{\cal G}}({\cal
G})}{dt}-f_{{\cal G}}({\cal G}){\cal G}+f({\cal G}),
\end{equation}

\begin{equation}\label{lambdat}
\lambda(t)=\dot{\phi}^{-2}\left(\frac{\rho}{2}+\dot{H}+4H(2\dot{H}-H^{2})\frac{df_{{\cal
G}}({\cal G})}{dt}+4H^{2}\frac{d^{2}f_{{\cal G}}({\cal
G})}{dt^{2}}\right)+\epsilon.
\end{equation}
Equation (\ref{SF}) follows easily by combining Eqs. (\ref{FGFR-1})
and (\ref{FGFR-2}). Therefore, for a given $f({\cal G})$ one can
specify the evolution of the scale factor (or the corresponding
Hubble parameter) and obtain the scalar field potential which is
responsible for this evolution. In addition, the inverse process is
easy to be realized, that is, for a given scalar potential and for
an arbitrarily given cosmological evolution, it is possible to
reconstruct the corresponding $f({\cal G})$ gravity. In the
following sections, we exemplify the two aforementioned
reconstruction methods.

\subsection{Some Examples for $f({\cal G})=A{\cal G}^{2}$ gravity and $\phi=\sqrt{U_{0}}t$ }

The simplest choice for the function $f({\cal G})$ is the following,
\begin{equation}\label{co1}
f({\cal G})=A{\cal G}^{2}.
\end{equation}

In this section, we shall assume that the scalar field $\phi$ is
related to the cosmic time $t$ as, $\phi=\sqrt{U_{0}}t$, which means
that $U(\phi)=U_{0}=\mbox{const}$. For the bounce case with scale
factor given in Eq. (\ref{sc1}) and Hubble rate given in Eq.
(\ref{hubra1}), the corresponding potential of the scalar field,
provided by Eq. (\ref{p1}), takes the following form,

\begin{equation}\label{app1}
V(t)=A\alpha^{5} U_{0}^{-1} \phi^{2}\sum_{n=0}^{3}C_{n}(\alpha
U_{0}^{-1} \phi^{2})^{n}+V_{0},
\end{equation}

where the parameter $V_0$, which is equal to,
$$
V_{0}=3(2 c)^{2}U_{0}^{-1} \phi^{2}+4 c+\epsilon U_{0}
$$
stands for the potential of the scalar field in the case of General
Relativity (which can be achieved if $A=0$), and $C_{n}$ are
numerical coefficients. For completeness, we provide the full
expression for the scalar potential in the Appendix. The
corresponding Lagrange multiplier function $\lambda (t)$, which is
given in Eq. (\ref{lambdat}), is equal to,
\begin{align}\label{mlabdat}
\lambda (\phi)= -\frac{36864 A c^5}{U_0}+\epsilon +\frac{2 c}{\phi
^2}+\frac{\rho }{\phi ^2}-\frac{221184 A c^6 \phi
^2}{U_0^2}+\frac{98304 A c^7 \phi ^4}{U_0^3}
\end{align}
In the same way, we may easily find the scalar potential and the
auxiliary function $\lambda (\phi)$, for the case of the superbounce
with scale factor and Hubble rate given in Eqs. (\ref{scsupe}) and
(\ref{hu}) respectively.  The corresponding scalar potential is,
\begin{align}\label{vpotsup}
& V (\phi)=\frac{12288 A \left(-12+44 c^2-75 c^4+28 c^6\right)
U_0^4}{c^{16} \left(-t_s \sqrt{U_0}+\phi \right)^8}\\
\notag & +\frac{c^{12} \sqrt{U_0} \left(-t_s \sqrt{U_0}+\phi
\right)^6 \left(c^4 t_s^2 U_0 \epsilon +c^4 \epsilon  \phi ^2-2
\sqrt{U_0} \left(-6+2 c^2+c^4 t_s \epsilon  \phi
\right)\right)}{c^{16} \left(-t_s \sqrt{U_0}+\phi \right)^8}
\end{align}
while the Lagrange multiplier function $\lambda (\phi)$ equals to,
\begin{align}\label{edgy}
& \lambda (\phi)=\frac{24576 A \left(-4-12 c^2+7 c^4\right)
U_0^4}{c^{14} \phi ^2 \left(-t_s \sqrt{U_0}+\phi \right)^8} \\
\notag & +\frac{c^{12} \left(-t_s \sqrt{U_0}+\phi \right)^6 \left(-2
c^2 t_s \sqrt{U_0} \phi  \left(\rho +\epsilon  \phi ^2\right)+c^2
\phi ^2 \left(\rho +\epsilon  \phi ^2\right)+U_0 \left(-2+c^2 t_s^2
\left(\rho +\epsilon  \phi ^2\right)\right)\right)}{c^{14} \phi ^2
\left(-t_s \sqrt{U_0}+\phi \right)^8}
\end{align}
Another example that produces Little Rip cosmology, is described by
the following Hubble rate,
$$
H(t)=H_{0}\exp(\alpha t), \quad \alpha>0.
$$
with the corresponding scalar potential being of the following form,
\begin{equation}
V(\phi)=A H^{5}_{0}\exp(5\gamma
\phi)\sum_{n=0}^{3}C_{n}\alpha^{3-n}H_{0}^{n}\exp(n\gamma
\phi)+V_{0},
\end{equation}
$$
V_{0}=3H_{0}^{2}\exp(2\gamma \phi)+2H_{0}\alpha\exp(\gamma
\phi)+\epsilon U,\quad \gamma=\alpha U^{-1/2}.
$$
In order to better understand what is the difference between the
Lagrange multiplier mimetic matter with a scalar potential, which
was studied in Ref. \cite{mukhanov2}, and Lagrange multiplier
$f(\cal{G})$ gravity, we shall study in detail some examples that
were presented in \cite{mukhanov2}. It is worth recalling some
features of Lagrange multiplier mimetic matter, with scalar
potential, and for a detailed presentation, the reader is referred
to \cite{mukhanov2}. The action of the mimetic matter with scalar
potential and with Lagrange multiplier, is given below \cite{viknew},
\begin{equation}\label{shamimts}
S=\int \mathrm{d}x^4 \left[ -\frac{1}{2}R+\lambda \left(g^{\mu
\nu}\partial_{\mu}\phi\partial_{\nu}\phi)-1\right)-V(\phi
)+L_m\right]
\end{equation}
with $L_m$ denoting as usual the matter Lagrangian. In the context
of the theory developed in \cite{mukhanov2}, the Lagrange multiplier
is equal to,
\begin{equation}\label{sed}
\lambda =\frac{1}{2}\left (G-T-4V(\phi)\right),
\end{equation}
with $G$ and $T$ standing for the trace of the Einstein tensor and
for the trace of the energy momentum tensor respectively. The form
of the Lagrange multiplier appearing in Eq. (\ref{sed}) is obviously
different from the Lagrange multiplier appearing in Eq.
(\ref{lambdat}), but one can easily verify that if we take
$U(\phi)=1$ and also $f({\cal{G}})=0$, these become identical. We
now investigate the new features that the $f(\cal{G})$ modification
brings along. We choose again the $f(\cal{G})$ gravity to be of the
form given in Eq. (\ref{co1}) and also we assume that
$\phi=\sqrt{U_{0}}t$. We shall start our comparison analysis, by
studying the inflaton case presented in \cite{mukhanov2}, where it
was shown that a cosmological evolution with scale factor equal to,
\begin{equation}\label{inflat}
a(t)= e^{-\sqrt{\frac{\alpha}{12}}t^2}
\end{equation}
with $\alpha$ a constant parameter. The potential that generates
this kind of evolution, always in the context of Ref.
\cite{mukhanov2}, is equal to,
\begin{equation}\label{potgener}
V(\phi )=\frac{\alpha \phi^2}{e^{\phi}+1}
\end{equation}
In contrast, in the case of the Lagrange multiplier $f(\cal{G})$
modified gravity, the potential that generates the cosmological
evolution described by Eq. (\ref{inflat}), is equal to,
\begin{equation}\label{segde}
V(\phi)= -\frac{2 \sqrt{\alpha }}{\sqrt{3}}+\sqrt{U_0} \epsilon
+\frac{\alpha  \phi ^2}{U_0}+\frac{256 A \alpha ^{5/2} \phi
^2}{\sqrt{3} U_0}-\frac{1088 A \alpha ^3 \phi ^4}{3
U_0^2}+\frac{1408 A \alpha ^{7/2} \phi ^6}{9 \sqrt{3}
U_0^3}-\frac{64 A \alpha ^4 \phi ^8}{9 U_0^4},
\end{equation}
which is clearly much different than the one in Eq.
(\ref{potgener}). In addition, the corresponding Lagrange multiplier
function $\lambda (\phi)$ is equal to,
\begin{equation}\label{lamdnewfndt}
\lambda (\phi )=\epsilon -\frac{\sqrt{\alpha }}{\sqrt{3} \phi
^2}+\frac{\rho }{\phi ^2}+\frac{128 A \alpha ^{5/2} \left(9 \sqrt{3}
U_0^2-27 U_0 \sqrt{\alpha } \phi ^2-2 \sqrt{3} \alpha \phi
^4\right)}{27 U_0^3}
\end{equation}
which is obviously different from the corresponding one in ordinary
Lagrange multiplier mimetic matter theory given in Eq. (\ref{sed}).
It is obvious that within the context of Lagrange multiplier
$f(\cal{G})$ modified theory of gravity, there are many ways of
generating various cosmological scenarios, since there is much
freedom in choosing the potential $U(\phi )$ and also the function
$f(\cal{G})$. In this way, almost any cosmological scenario can be
generated, by suitably choosing the two aforementioned functions.
Before closing this section, we shall study another example
presented in \cite{mukhanov2}, which describes mimetic matter as
quintessence. The scale factor of the corresponding cosmological
evolution, is equal to \cite{mukhanov2},
\begin{equation}\label{qiuint}
a(t)=t^{\frac{2}{3(1+w)}}
\end{equation}
which actually describes the behavior of mimetic matter when the
Universe is dominated by some other matter fluid present, with
equation of state $p=w \rho$. The potential that generates the
cosmological evolution (\ref{qiuint}), is equal to \cite{mukhanov2},
\begin{equation}\label{quintpot}
V(\phi )\simeq \frac{\alpha}{t^2}
\end{equation}
where $\alpha$ is some arbitrary constant. In the case of Lagrange
multiplier $f(\cal{G})$ gravity, the potential that can generate the
cosmological evolution of Eq. (\ref{qiuint}), is equal to,
\begin{equation}\label{cosmoevo}
V(\phi )=\frac{U_0^4 \left(4096 A (1+3 w) (67+w (149+84
w))-\frac{972 w (1+w)^6 \phi ^6}{U_0^3}+\frac{729 (1+w)^8 \epsilon
\phi ^8}{U_0^{7/2}}\right)}{729 (1+w)^8 \phi ^8},
\end{equation}
and the corresponding Lagrange multiplier function $\lambda (\phi
)$, is equal to,
\begin{equation}\label{lamquint}
\lambda (\phi )=\epsilon +\frac{8192 A U_0^4 (1+3 w) (23+21 w)}{729
(1+w)^7 \phi ^{10}}-\frac{2 U_0}{3 (1+w) \phi ^4}+\frac{\rho }{\phi
^2}.
\end{equation}
which are clearly different from the ones given in Eqs.
(\ref{quintpot}) and (\ref{sed}).

As we already mentioned, in the context of Lagrange multiplier
$f({\cal G})$ modified gravity, there is much freedom in
reconstructing the scalar potential for the scalar field, since the
function $U(\phi)$ and the function $f({\cal G})$ itself can be
arbitrarily chosen. Let us consider as a final example the case for
which the function that relates the scalar field $\phi$ to the
cosmic time $t$ is of the following form, $\phi= \sqrt{U_{0}}\ln t$,
so the function $U(\phi)$ is of the following form,
\begin{equation}\label{r}
U(\phi)=U_{0}\exp(-2U_{0}^{-1/2}\phi).
\end{equation}
Assuming that,
$$
H_{0}=\alpha t
$$
and also that the function $f({\cal G})$ is again $f({\cal
G})=A{\cal G}^{2}$, the corresponding scalar field potential
$V(\phi)$, is of the form,
\begin{equation}
V(\phi)=A\alpha^5 \exp(2U_{0}^{-1/2}
\phi)\sum_{n=0}^{3}C_{n}\alpha^{n}\exp(nU_{0}^{-1/2}\phi)+V_{0},
\end{equation}
where $V_0$ in this case stands for,
$$
V_{0}=2\alpha+3\alpha^2\exp(2U_{0}^{-1/2}\phi)+\epsilon
U_{0}\exp(-2U_{0}^{-1/2}\phi).
$$
In the same way, more elaborated universe evolution unifying
inflation, dark energy and dark matter maybe presented.

\subsection{Reconstruction of $f({\cal G})$ gravity given the Lagrange multiplier function and the scale factor}

In this section we consider the inverse reconstruction method, which
enables us to find the Lagrange multiplier $f({\cal G})$ modified
gravity, given the scale factor of the cosmological evolution and
the Lagrange multiplier function. Let us explicitly demonstrate how
this method works. We start off with Eq. (\ref{lambdat}), which can
be rewritten as:
\begin{equation}
(\lambda(t)-\epsilon)\dot{\phi}^{2}-\frac{\rho}{2}-\dot{H}-4a(t)\frac{d}{dt}\left(\frac{H^{2}}{a(t)}\frac{df_{{\cal
G}}}{dt}\right)=0.
\end{equation}
Solving this equation with respect to $f_{{\cal G}}(t)$ we have:
\begin{equation}\label{above}
f_{{\cal G}}(t)=\frac{1}{4}\int
dt_{1}\frac{a(t_{1})}{H^{2}(t_{1})}\int^{t_{1}}
\frac{dt_{2}}{a(t_{2})}\left((\lambda(t_{2})-\epsilon)\dot{\phi}^{2}(t_{2})-\dot{H}(t_{2})-\frac{\rho_{0}}{2a^{3}(t_{2})}\right).
\end{equation}
Note that for $\lambda=1$, Eq. (\ref{above}) in fact coincides with
(\ref{SolG}) (with $C=0$) for $g(t)=df_{{\cal G}}/dt$, as was
expected. The potential of scalar field as a function of time is
cast as follows,
\begin{equation}\label{potentnewder}
V(t)=(2\lambda
(t)-\epsilon)\dot{\phi}^{2}(t)+3H^{2}(t)-\rho(t)+6a(t)H(t)\int\frac{dt_{1}}{a(t_{1})}\left((\lambda(t_{1})-\epsilon)\dot{\phi}^{2}(t_{1})-\dot{H}(t_{1})-\frac{\rho_{0}}{2a^{3}(t_{1})}\right)
\end{equation}
$$
-f_{{\cal G}}(t){\cal G}(t)+f(t).
$$
We assume that the function that relates the scalar field $\phi$ and
the cosmic time $t$ is denoted as $t=y(\phi)$. We also write the
scale factor $a(t)$ in the following form,
\begin{equation}\label{scfexp}
a(t)=\exp(h(t)).
\end{equation}
Then, the Hubble parameter is simply written in terms of $h(t)$, as
$H(t)=\dot{h}(t)$. Considering for simplicity the case that no
matter fluids are present, that is $\rho=0$, the potential of the
scalar field given in Eq. (\ref{potentnewder}), becomes,
\begin{equation}\label{sncstax}
V(\phi)=3\dot{h}^{2}(y(\phi))-\frac{\epsilon}{y'^{2}(\phi)}-6\mbox{e}^{h(y(\phi)}\dot{h}(y(\phi))\int
{d\phi_{1}}y'(\phi_{1}){\mbox{e}^{-h(y(\phi_{1}))}}\left(\frac{\epsilon}{y'^{2}(\phi_{1})}+\ddot{h}(y(\phi_{1}))\right)+
\end{equation}
$$
+\frac{2\lambda(\phi)}{y'^{2}(\phi)}+6\mbox{e}^{h(y(\phi))}\dot{h}(y(\phi))\int
{d\phi_{1}}{\mbox{e}^{-h(y(\phi_{1}))}}\left(\frac{\lambda(\phi_{1})}{y'(\phi_{1})}\right)-
$$
$$
-f_{{\cal G}}(\phi){\cal G}(y(\phi))+\int d\phi_{1} f_{{\cal
G}}(\phi_{1})y'(\phi_{1}){{\cal G}_{y}}(y(\phi_{1}))
$$
and therefore, the function $f_{{\cal G}}$ as a function of the
scalar field is equal to,
\begin{equation}\label{centreq}
f_{{\cal G}}(\phi)=-\frac{1}{4}\int d\phi_{1}
{\dot{h}^{-2}(y(\phi_{1}))\mbox{e}^{h(y(\phi_{1}))}}{y'(\phi_{1})}\int^{\phi_{1}}
{d\phi_{2}}y'(\phi_{2}){\mbox{e}^{-h(y(\phi_{2}))}}\left(\frac{\epsilon}{y'^{2}(\phi_{2})}+\ddot{h}(y(\phi_{2}))\right)+
\end{equation}
$$
+\frac{1}{4}\int d\phi_{1}
{\dot{h}^{-2}(y(\phi_{1}))\mbox{e}^{h(y(\phi_{1}))}}{y'(\phi_{1})}\int^{\phi_{1}}
{d\phi_{2}}{\mbox{e}^{-h(y(\phi_{2}))}}\left(\frac{\lambda(\phi_{2})}{y'(\phi_{2})}\right).
$$
Then, for a given cosmological evolution, in terms of some specific
scale factor, by choosing the arbitrary Lagrange multiplier function
$\lambda(\phi)$, we can easily obtain the corresponding $f({\cal
G})$ gravity, and also finally find the resulting form of the
potential $V(\phi)$, which we used earlier for the derivation of the
$f({\cal G})$ gravity.

Let us exemplify how this inverse reconstruction method works, by
using some illustrative examples. For simplicity, we shall assume
that the field $\phi$ is related to the cosmic time as,
$\phi=\sqrt{U_{0}} t$, so we define the function $y(\phi )$ to be
the following,
\begin{equation}\label{newdef}
y(\phi)=U_{0}^{-1/2}\phi
\end{equation}
which we extensively use in the following considerations. Of course,
one can in principle choose a different function, but this will only
perplex the equations, without changing the qualitative feature of
our reconstruction method.

We start off with the superbounce, with scale factor and Hubble rate
appearing in Eqs. (\ref{scsupe}) and (\ref{hu}) respectively. The
function $h(t)$ is in this case equal to,
\begin{equation}\label{htfn1}
h(t)=\ln \left( -t+t_s\right )^{\frac{2}{c^2}}
\end{equation}
In addition, we take the function $\lambda (t)$, to be equal to,
\begin{equation}\label{lamdgaw}
\lambda (t)= c t
\end{equation}
Then, by combining Eqs. (\ref{centreq}), (\ref{newdef}),
(\ref{htfn1}) and (\ref{lamdgaw}), we obtain the function $f_{{\cal
G}}(\phi)$, which is equal to,
\begin{align}\label{fglysi1}
& f_{{\cal G}}(\phi)= \frac{c^7 \left(-\frac{c^2 t_s^4 \phi
}{\sqrt{U_0}}+\frac{\left(1+c^2\right) t_s^3 \phi
^2}{\sqrt{U_0}}-\frac{2 t_s^2 \phi
^3}{\sqrt{U_0}}-\frac{\left(-3+c^2\right) t_s \phi ^4}{2
\sqrt{U_0}}+\frac{\left(-2+c^2\right) \phi ^5}{5
\sqrt{U_0}}\right)}{32 \left(2-3 c^2+c^4\right) U_0}
\\ \notag & -\frac{c^4 \phi  (-2 t_s+\phi ) \left(2 c^2 \left(2+c^2\right) t_s^2 \epsilon
 -2 c^2 \left(2+c^2\right) t_s \epsilon  \phi +U_0 \left(-8+\frac{c^4 \epsilon \phi ^2}{U_0}
+2 c^2 \left(2+\frac{\epsilon  \phi
^2}{U_0}\right)\right)\right)}{64 \left(-4+c^4\right) U_0}
\end{align}
Then, the $f(\cal {G})$ gravity easily follows, by using the
expression for the Gauss-Bonnet invariant, which for the case of the
superbounce is given in Eq. (\ref{gabne}), by integrating with
respect to $\cal{G}$. The final expression is quite large and for
simplicity reasons we give it in the Appendix. In the same way, by
using Eqs. (\ref{sncstax}), (\ref{newdef}), (\ref{htfn1}) and
(\ref{lamdgaw}), we obtain the potential of the scalar field, which
is,
\begin{align}\label{supotnew1}
& V(\phi )= -\frac{\epsilon }{U_0}+\frac{12}{c^4 (t_s-\phi
)^2}+\frac{2 c \phi }{U_0^{3/2}}-\frac{6 c \left(2 t_s \phi -2 \phi
^2+c^2 \left(-t_s^2+\phi ^2\right)\right)}{\left(2-3 c^2+c^4\right)
U_0^{3/2} (t_s-\phi )} \\ \notag & -\frac{12 \left(c^2
\left(2+c^2\right) t_s^2 \epsilon -2 c^2 \left(2+c^2\right) t_s
\epsilon  \phi +U_0 \left(-4+\frac{c^4 \epsilon \phi ^2}{U_0}+2 c^2
\left(1+\frac{\epsilon  \phi ^2}{U_0}\right)\right)\right)}{c^2
\left(-4+c^4\right) U_0 (t_s-\phi )^2} \\ \notag & + \frac{96
\left(\frac{4}{c^4 (t_s-\phi )^2}-\frac{2}{c^2 (t_s-\phi
)^2}\right)\left(c^7 \left(-\frac{c^2 t_s^4 \phi
}{\sqrt{U_0}}+\frac{\left(1+c^2\right) t_s^3 \phi
^2}{\sqrt{U_0}}-\frac{2 t_s^2 \phi
^3}{\sqrt{U_0}}-\frac{\left(-3+c^2\right) t_s \phi ^4}{2
\sqrt{U_0}}+\frac{\left(-2+c^2\right) \phi ^5}{5
\sqrt{U_0}}\right)\right)}{c^4 (t_s-\phi )^232 \left(2-3
c^2+c^4\right) U_0} \\ \notag & -\frac{96 \left(\frac{4}{c^4
(t_s-\phi )^2}-\frac{2}{c^2 (t_s-\phi )^2}\right)\left(c^4 \phi  (-2
t_s+\phi ) \left(2 c^2 \left(2+c^2\right) t_s^2 \epsilon -2 c^2
\left(2+c^2\right) t_s \epsilon \phi +U_0 \left(-8+\frac{c^4
\epsilon \phi ^2}{U_0}+2 c^2 \left(2+\frac{\epsilon  \phi
^2}{U_0}\right)\right)\right)\right)}{c^4 (t_s-\phi )^264
\left(-4+c^4\right) U_0} \\ \notag & -\frac{3 \left(-2+c^2\right)
\left(t_s^2 \left(-2 c^3 t_s^3+5 c^5 t_s^3+3 c^7 t_s^3-40
U_0^{3/2}-5 c^6 t_s^2 \sqrt{U_0} \epsilon -5 c^4 \sqrt{U_0} \left(4
U_0+t_s^2 \epsilon \right)+10 c^2 \sqrt{U_0} \left(6 U_0+t_s^2
\epsilon \right)\right)\right)}{5 c^4
U_0^{3/2} \left(4-4 c^2-c^4+c^6\right) (t_s-\phi )^4} \\
\notag & +\frac{+\frac{40 U_0^{3/2}}{\left(2+c^2\right) (t_s-\phi
)^2}+\frac{8 c^3 \phi }{-1+c^2}+\frac{20 c^2 \left(c t_s-\sqrt{U_0}
\epsilon \right) \ln [-t_s+\phi ]}{-2+c^2}}{5 c^4 U_0^{3/2}}
\end{align}
In the same way, we can easily obtain the $f(\cal {G})$ gravity for
the case of the bounce with scale factor (\ref{sc1}). Using the same
conventions as in the case of the superbounce, the resulting
expression for the $f(\cal {G})$ gravity is,
\begin{align}\label{alignher}
& f({\cal {G}})=-\frac{1}{4 U_0^2}c
\sqrt{-\frac{4}{c}+\frac{\sqrt{16 c^6-\frac{2 c^4 G}{3}}}{c^4}}
\\ \notag & \times
\left(8+\frac{\sqrt{16 c^6-\frac{2 c^4 G}{3}}}{c^3}+\frac{3}{16} c
\left(-\frac{4}{c}+\frac{\sqrt{16 c^6-\frac{2 c^4
G}{3}}}{c^4}\right)^{3/2} \sqrt{\pi } U_0 (2 c U_0+\epsilon )\right
)\\ \notag & \times G^{2 2}_{3 4}  \left(\begin{array}{c}
                                                      -1,0,1 \\
                                                      0,0,-2,-\frac{1}{2} \\
                                                    \end{array}\Big{|}\frac{12 c^3 U_0-\sqrt{144 c^6-6 c^4 G} U_0}{48 c^3}
                                                  \right) \\ \notag
                                                  & +\frac{1}{4} \sqrt{\frac{-36 c^3 \pi +3 \sqrt{144 c^6-6 c^4 G} \pi }{c^4}} U_0 (2 c U_0+\epsilon
                                                  )G^{2 2}_{3 4} \left(\begin{array}{c}
                                                      0,0,1 \\
                                                      0,0,-1,-\frac{1}{2} \\
                                                    \end{array}\Big{|}\frac{12 c^3 U_0-\sqrt{144 c^6-6 c^4 G} U_0}{48 c^3}
                                                  \right)\end{align}
with $G_{p q}^{m n}$, the Meijer $G$-function. The resulting
expression for the scalar potential is too complicated, so we
provide the details for it in the Appendix.

As a final example, we shall consider a non-monotonic evolution for
the Hubble rate which has the following form,
$$
H(t)=\frac{2}{t}+\frac{2}{t_{f}-t}.
$$
Notice that a Big Rip singularity occurs at $t=t_{f}$. It easily
follows that,
$$
h(\phi)=\ln\left(\frac{\phi}{U_{0}}\right)-\ln\left(\frac{(\phi_{f}-\phi)^{2}}{U_{0}}\right),
$$
$$
\dot{h}(\phi)=\frac{2\sqrt{U_{0}}}{\phi}+\frac{2\sqrt{U_{0}}}{\phi_{f}-\phi},\quad
\ddot{h}(\phi)=-\frac{2U_{0}}{\phi^{2}}+\frac{2U_{0}}{(\phi_{f}-\phi)^{2}},
\quad \phi_{f}=\sqrt{U_{0}}t_{f}.
$$
Without getting into too much detail, the function $f_{{\cal
G}}(\phi)$ is equal to,
\begin{equation}
f_{{\cal
G}}(\phi)=-\frac{1}{16U_{0}\phi_{f}^{2}}\left(\frac{\epsilon\phi^{6}}{6}+C_{1}\phi^5+\frac{2\epsilon\phi_{f}}{25}\phi^{5}-
\frac{2\epsilon\phi_{f}}{5}\phi^{5}\ln\phi-\frac{\epsilon\phi_{f}^{2}}{4}\phi^{4}-\frac{2\phi_{f}}{3}\phi^{3}+
\frac{\phi_{f}^{2}}{2}\phi^{2}+C_{2}\right)+
 \end{equation}
$$
+\frac{1}{4}\int d\phi_{1}
\frac{\dot{h}^{-2}(y(\phi_{1}))\mbox{e}^{h(y(\phi_{1}))}}{y'^{2}(\phi_{1})}\int^{\phi_{1}}
{d\phi_{2}}{\mbox{e}^{-h(y(\phi_{2}))}}\left(\frac{\lambda(\phi_{2})}{y'(\phi_{2})}\right).
$$
while the corresponding scalar field potential is equal to,
\begin{equation}
V(\phi)=\frac{U_{0}}{(\phi_{f}-\phi)^{3}}\left(4\phi^{3}_{f}\phi^{-2}+12\phi^{2}_{f}\phi^{-1}+\epsilon(11\phi^{3}_{f}+3\phi^{2}_{f}\phi-15\phi_{f}\phi^{2}+\phi^{3}+
24\phi^{2}_{f}\phi\ln\phi)-60C_{1}\phi_{f}\phi\right)+
\end{equation}
$$
+\frac{2\lambda(\phi)}{y'^{2}(\phi)}+6\mbox{e}^{h(y(\phi))}\dot{h}(y(\phi))\int
{d\phi_{1}}{\mbox{e}^{-h(y(\phi_{1}))}}\left(\frac{\lambda(\phi_{1})}{y'(\phi_{1})}\right)-
$$
$$
-\int d\phi_{1} \frac{f_{{\cal G}}(\phi_{1})}{d\phi_{1}}{{\cal
G}}(y(\phi_{1})).
$$
Notice that the last integral can be evaluated analytically, but we
omit the resulting expression for the sake of brevity. In
conclusion, with this inverse reconstruction method, if the Lagrange
multiplier function $\lambda(\phi)$ and the scale factor $a(t)$ are
specified, the corresponding Lagrange multiplier $f({\cal G})$
gravity follows, as we explicitly demonstrated in the previous
examples. Of course, it is to be understood that if the integrals
cannot be evaluated analytically, further assumptions should be
taken into account, but the general theoretical framework is the
same.

\section{Discussion}

It is worth discussing certain issues related to the mimetic $f({\cal G})$ gravity models we studied in this paper. In particular one possible question that comes to mind when dealing with this kind of theories is the following: all the models we constructed in the present study involve an $f({\cal G})$ function, with ${\cal G}$ being the Gauss-Bonnet term which in the usual case is harmless, since it can be integrated out topologically. Now the question is if the theory constructed by using a function of the Gauss-Bonnet invariant, contains any extra degrees of freedom or not. In our case however, note that the $f({\cal G})$ gravity is considered here as an effective string-inspired gravity theory. Due to non-linear functional dependence on G, the function $f({\cal G})$ is not a topological invariant anymore. The introduction of the $f({\cal G})$ term, introduces to the theory an effective scalar degree of freedom, as in many cases the model maybe rewritten as one containing a string-inspired $V(\phi)\mathcal{G}$ term. In addition, in the context of the mimetic $f({\cal G})$ gravity, there is an extra scalar degree of freedom, that of the scalar field $\phi$, which accounts for the dark matter content of the theory. Now we come to another important issue that we need to briefly discuss. As we demonstrated in this paper, the mimetic $f({\cal G})$ cosmological scenario can serve as a theoretical framework that at the same time can describe inflation, dark energy and even dark matter. Although it is quite appealing to have a geometric description of the cosmic history of our Universe, an important question rises, with regards to the inflationary particle content of the theory. In particular,  for inflation, the most important and successful achievement was to predict a nearly scale invariant power spectrum which is verified by CMB experiments in the past decade. However, this achievement is based on the fundamental assumption that inflation is realized by a slow-rolling scalar field in the early universe. Without such a scalar field, how can the CMB spectrum be realized  in modified gravity models?

This question is very interesting and we shall try to briefly address this, so let us first consider the standard mimetic gravity, presented in \cite{mukhanov1,mukhanov2}, where the authors considered the standard Einstein-Hilbert mimetic extension. As it was shown in \cite{mukhanov2}, the metric perturbation is related to the higher derivative of the scalar field's linear perturbation $\dot{\delta \phi}$, and in effect, the resulting picture is that, the perturbations
behave as dust, with the speed of sound being zero, even for mimetic matter. Therefore, the quantum fluctuations of mimetic matter are ill-defined, and therefore the large scale structure cannot be explained by the minimal Einstein-Hilbert mimetic matter theory. An extra scalar field is required to play the role of the curvaton or explicitly modify the mimetic dark matter model, in order for the large scale structure being consistently described. Therefore, the mimetic $f({\cal G})$ could potentially provide such an extension, by offering extra degrees of freedom to the theory. The corresponding extra scalar degree of freedom we discussed earlier, may play such a role in the resulting theory. In \cite{mukhanov2}, the authors modified by hand the minimally extended mimetic Einstein-Hilbert Lagrangian, thus being able to produce a red-tilted spectral index. It would be therefore quite interesting to investigate whether this extra scalar term can originate by an $f({\cal G})$ theory or from any other mimetic extension of modified gravity. However, this task exceeds the purposes of this paper and we hope to address such issues in the future.

\section{Conclusions \label{SecVIII}}

In this paper we considered the generalization of ordinary Jordan
frame $f({\cal G})$ gravity, to mimetic $f({\cal G})$ gravity. In
this new theory, in principle it is possible to resolve two
important cosmological problems, firstly to obtain the Universe's
accelerated expansion without the need to introduce extra fields
(inflaton, dark energy scalar) and secondly to provide an answer to
the problem of dark matter. Specifically, within the context of
mimetic $f({\cal G})$ gravity, the cosmological equations of motion
are almost the same with these that result in ordinary Jordan frame
$f({\cal G})$ gravity, with the only difference being that the
contribution of mimetic matter appears in the mimetic $f({\cal G})$
gravity case. After having provided the formalism of mimetic
$f({\cal G})$ gravity, we were able to realize various cosmological
evolutions with mimetic $f({\cal G})$ gravity, and highlighted the
difference between ordinary and mimetic $f({\cal G})$ gravity. In
principle, in the case of mimetic $f({\cal G})$ gravity, there is
much more freedom in providing the $f({\cal G})$ gravity that can
realize a certain cosmological evolution. This can be one of the
attributes of the theory, since a quite large number of cosmological
evolutions can be successfully realized in the context of mimetic
$f({\cal G})$ gravity. We also compared the resulting mimetic
$f({\cal G})$ gravities, to standard mimetic gravity models that
exist in the literature.

We also modified the ordinary Jordan frame $f({\cal G})$ gravity, to
include a Lagrange multiplier term. After describing in detail how
the cosmological equations are modified within this new formalism,
we provided two quite general reconstruction techniques, which in
principle can be useful for the realization of various cosmological
models. Particularly, it is possible in the context of Lagrange
multiplier $f({\cal G})$ gravity, given the cosmological evolution
and the $f({\cal G})$ gravity, to find the scalar field potential
$V(\phi )$ and the Lagrange multiplier function $\lambda (t)$. In
addition, the inverse procedure is possible, that is, given the
cosmological evolution and the Lagrange multiplier function $\lambda
(t)$, it is possible to reconstruct the corresponding $f({\cal G})$
gravity. We supported our theoretical considerations by using some
illustrative examples, for which we applied both reconstruction
methods.

As a general remark, we must note that in the context of both
mimetic $f({\cal G})$ gravity and Lagrange multiplier $f({\cal G})$
gravity, there exist a large class of $f({\cal G})$ gravities that
can realize a specific cosmological evolution. We believe that this
feature is an attribute of the theory, since it is possible, by
suitably choosing the parameters and functions, to get analytic
results. This was not always possible in ordinary Jordan frame
$f({\cal G})$ gravity.

An issue we did not address in this paper is the
realization of singular cosmology, in the context of mimetic
$f({\cal G})$ gravity or Lagrange multiplier $f({\cal G})$ gravity.
By singular cosmology we mean the appearance of finite time
singularities \cite{Nojiri:2005sx} (with the Big Rip  \cite{BigRip}
being the most elaborated one) in the cosmological evolution. Among
all the finite time singularities, the most interesting ones are the
non-crushing types singularities, for example, Type II \cite{Barrow}
or the mildest among them, the Type IV \cite{Nojiri:2005sx,pap1}.
Since these singularities can be consistently incorporated in
scalar-tensor theories \cite{pap1}, it would be interesting to
realize such singular cosmological evolution from mimetic $f({\cal
G})$ and from Lagrange multiplier $f({\cal G})$, and compare the
results to ordinary $f({\cal G})$ gravity. Special emphasis for this
task, should be given near the Type IV singularity of course. Work
is in progress towards this direction and we hope to report on this
issue soon.

Finally, let us discuss another issue having to do with the possible observational distinction between the various modified gravity theories. Indeed, these theories can in principle be very degenerate with regards to their observational prediction so the fundamental question that comes to mind is, whether there is any possible way of distinguishing these theories at a pragmatic level. One good idea is to these these modified gravity models by using weak lensing data, owing to the fact that the trajectories of light may differ, depending on different geometric terms. For some relevant works on this issue see for example \cite{refe1}. In addition, the growth index may provide some useful insights on that respect, see for example \cite{refe2}.

\section*{Acknowledgments}

This work is supported in part by project 14-02-31100 (RFBR, Russia)
and project 2058 (MES, Russia) (AVA),  by MINECO (Spain), projects
FIS2010-15640 and FIS2013-44881 (SDO).

\section*{Appendix: Resulting expressions for $f({\cal {G}})$ gravities and for scalar potentials}

Here we quote the complicated expressions we mentioned in the main
text of this paper. The full expression for the potential appearing
in Eq. (\ref{app1}) is equal to,
\begin{equation}\label{eqnappend}
V(\phi )=4 c+12 c^2 t^2-73728 A c^5 t^2-626688 A c^6 t^4-540672 A
c^7 t^6-147456 A c^8 t^8+\sqrt{U_0} \epsilon
\end{equation}
In addition, by integrating Eq. (\ref{fglysi1}), with respect to
$\cal{G}$, we obtain the $f({\cal{G}})$ function, which is equal to,
\begin{align}\label{fnexprforfg}
& f(\phi ({\cal{G}}))= -\frac{1}{c^4 U_0}12 \Big{(}-2+c^2\Big{)}
\Big{(}\frac{2 c^3 U_0^{3/2} \phi ({\cal{G}}) }{5
\Big{(}-1+c^2\Big{)}}-\frac{1}{20 \Big{(}4-4 c^2-c^4+c^6\Big{)}
\Big{(}-t_s+\frac{\phi ({\cal{G}}) }{\sqrt{U_0}}\Big{)}^4}
\\ \notag & \times t_s^2 \Big{(}2 c^3 t_s^3
\sqrt{U_0} \Big{(}10-20 \sqrt{U_0}+15 U_0-4
U_0^{3/2}\Big{)}+5 c^5 t_s^3 \Big{(}-4+6 \sqrt{U_0} \\
\notag &-4 U_0+U_0^{3/2}\Big{)}-40 \Big{(}-2
U_0^{3/2}+U_0^2\Big{)}+c^7 t_s^3 \Big{(}-10+10 \sqrt{U_0}-5
U_0^{3/2}+2 U_0^2\Big{)}
\\ \notag &-5 c^6 t_s^2 \sqrt{U_0} \Big{(}-4+6
\sqrt{U_0}-4 U_0+U_0^{3/2}\Big{)} \epsilon -5 c^4 \Big{(}-2
\sqrt{U_0}+U_0\Big{)} \Big{(}2 t_s^2 \epsilon -2 t_s^2 \sqrt{U_0}
\epsilon +U_0 \Big{(}4+t_s^2 \epsilon \Big{)}\Big{)}
\\ \notag &+10 c^2 \Big{(}-2
\sqrt{U_0}+U_0\Big{)} \Big{(}2 t_s^2 \epsilon -2 t_s^2 \sqrt{U_0}
\epsilon +U_0 \Big{(}6+t_s^2 \epsilon \Big{)}\Big{)}\Big{)}
\\ \notag & -\frac{1}{3
\Big{(}4-4 c^2-c^4+c^6\Big{)} \Big{(}-t_s+\frac{\phi ({\cal{G}})
}{\sqrt{U_0}}\Big{)}^3}2 t_s \Big{(}-1+\sqrt{U_0}\Big{)} \Big{(}2
c^5 t_s^3 \Big{(}-1+\sqrt{U_0}\Big{)}^2+c^7 t_s^3
\Big{(}-1+\sqrt{U_0}\Big{)}^2 \Big{(}1+\sqrt{U_0}\Big{)}
\\ \notag &-4 c^3 t_s^3
\Big{(}-1+\sqrt{U_0}\Big{)}^2 \sqrt{U_0}-8 U_0^{3/2}-2 c^6 t_s^2
\Big{(}-1+\sqrt{U_0}\Big{)}^2 \sqrt{U_0} \epsilon
\\ \notag & -2 c^4
\sqrt{U_0} \Big{(}t_s^2 \epsilon -2 t_s^2 \sqrt{U_0} \epsilon +U_0
\Big{(}2+t_s^2 \epsilon \Big{)}\Big{)}+4 c^2 \sqrt{U_0} \Big{(}t_s^2
\epsilon -2 t_s^2 \sqrt{U_0} \epsilon +U_0 \Big{(}3+t_s^2 \epsilon
\Big{)}\Big{)}\Big{)}
\\ \notag &+\frac{1}{\Big{(}4-4 c^2-c^4+c^6\Big{)}
\Big{(}-t_s+\frac{\phi ({\cal{G}}) }{\sqrt{U_0}}\Big{)}^2}\Big{(}-3
c^5 t_s^3 \Big{(}-1+\sqrt{U_0}\Big{)}^2 \sqrt{U_0}+2 c^3 t_s^3
\Big{(}-1+\sqrt{U_0}\Big{)}^2\\ \notag & \times \Big{(}-1+4
\sqrt{U_0}\Big{)} \sqrt{U_0}+4 U_0^2-c^7 t_s^3 \sqrt{U_0} \Big{(}1-3
U_0+2 U_0^{3/2}\Big{)}+3 c^6 t_s^2 \Big{(}-1+\sqrt{U_0}\Big{)}^2 U_0
\epsilon
\\ \notag & -6 c^2
U_0 \Big{(}U_0+t_s^2 \epsilon -2 t_s^2 \sqrt{U_0} \epsilon +t_s^2
U_0 \epsilon \Big{)}+c^4 U_0 \Big{(}3 t_s^2 \epsilon -6 t_s^2
\sqrt{U_0} \epsilon +U_0 \Big{(}2+3 t_s^2 \epsilon
\Big{)}\Big{)}\Big{)}
\\ \notag & -\frac{4 c^2 t_s
\Big{(}-1+\sqrt{U_0}\Big{)} U_0 \Big{(}c \Big{(}t_s-2 t_s
\sqrt{U_0}\Big{)}+c^3 t_s \sqrt{U_0}+\sqrt{U_0} \epsilon -c^2
\sqrt{U_0} \epsilon \Big{)}}{\Big{(}2-3 c^2+c^4\Big{)}
\Big{(}-t_s+\frac{\phi ({\cal{G}}) }{\sqrt{U_0}}\Big{)}}
\\ \notag & +\frac{c^2
U_0^{3/2} \Big{(}c t_s \Big{(}3-4 \sqrt{U_0}\Big{)}+c^3 t_s
\Big{(}-1+2 \sqrt{U_0}\Big{)}+\sqrt{U_0} \epsilon -c^2 \sqrt{U_0}
\epsilon \Big{)} \ln\left[-t_s+\frac{\phi ({\cal{G}})
}{\sqrt{U_0}}\right]}{2-3 c^2+c^4}\Big{)}
\end{align}
where $\phi ({\cal {G}})$ stands for,
\begin{equation}\label{name}
\phi ({\cal {G}})=\frac{2 \sqrt{2} 3^{1/4} c^4 \sqrt{G} \left(c^8
\left(-2+c^2\right) G\right)^{1/4}+c^8 G t_s}{c^8 G\text{
}\left(\sqrt{U_0}\right)^{-1}},{\,}{\,}{\,}
\end{equation}
or equivalently,
\begin{equation}\label{name1}
\phi ({\cal {G}})=\frac{-2 \sqrt{2} 3^{1/4} c^4 \sqrt{G} \left(c^8
\left(-2+c^2\right) G\right)^{1/4}+c^8 G t_s}{c^8 G
\left(\sqrt{U_0}\right)^{-1}}
\end{equation}
In conclusion, we have two different types of $f(\phi ({\cal{G}}))$
gravity, which can be found by substituting the two different forms
of $\phi ({\cal {G}})$ to Eq. (\ref{fnexprforfg}). Finally, the
corresponding potential for the $f({\cal{G}})$ gravity of Eq.
(\ref{alignher}), which corresponds to the bounce (\ref{sc1}), is
equal to,
\begin{align}\label{finapotnerbounce}
& V(\phi )=-\frac{\epsilon }{U_0}-\frac{4 c \phi }{U_0^{3/2}}+12 c^2
\phi ^2-\frac{6 \sqrt{c} e^{c \phi ^2} \sqrt{\pi } (2 c U_0+\epsilon
) \phi  \text{erf}\left[\sqrt{c} \phi \right]}{U_0} \\ \notag & +96
c^2 \phi ^2 \left(2 c+4 c^2 \phi ^2\right) \left(\frac{1}{32 c^2
U_0^{3/2} \phi }+\frac{\sqrt{\pi } (2 c U_0+\epsilon ) G^{2 1}_{2 3}
\left(\begin{array}{c}
                                                      0,1 \\
                                                      0,0,-\frac{1}{2} \\
                                                    \end{array}\Big{|}-c \phi ^2
                                                  \right)}{64 c^2
U_0}\right)\\ \notag & -\frac{c \phi  \left(4 \left(3+4 c \phi
^2\right)+12 c \sqrt{\pi } \sqrt{U_0} (2 c U_0+\epsilon ) \phi ^3
G^{2 2}_{3 4} \left(\begin{array}{c}
                                                     -1, 0,1 \\
                                                      0,0,-2,-\frac{1}{2} \\
                                                    \end{array}\Big{|}-c \phi ^2
                                                  \right)\right)}{U_0^{3/2}}\\
                                                  \notag &
                                                  +\frac{c \phi  \left(3 \sqrt{\pi } \sqrt{U_0} (2 c U_0+\epsilon )
                                                  \phi   G^{2 1}_{2 3}
\left(\begin{array}{c}
                                                      0,1 \\
                                                      0,0,-\frac{1}{2} \\
                                                    \end{array}\Big{|}-c \phi ^2
                                                  \right)\right)}{U_0^{3/2}}\end{align}
where $G^{m n}_{p q}$ is the Meijer G-function and $\mathrm{erf}(x)$
is the error function.

\end{document}